\documentclass[3p,twocolumn,10pt]{elsarticle} 
\pdfoutput=1
\usepackage{color}
\newcommand{\kstarkstar}{\Bs \to \Kstarz\!\Kstarzb}
\newcommand{\Bdkstarkstar}{\Bd \to \Kstarz \!\Kstarzb}
\newcommand{\BJpsiX}{\ensuremath{B\to J/\psi X}\xspace}
\newcommand{\jpsikstar}{\Bd \to \jpsi\!\Kstarz}

\usepackage{amsmath}

\usepackage{xspace}

\usepackage{epsfig}

\usepackage{rotating}

\usepackage{dcolumn}

\usepackage{url}

\usepackage{hyperref}

\usepackage{ifthen} 
\usepackage{subfigure}
\usepackage{rotating}

\usepackage{lineno}  
\usepackage{graphicx}  

\usepackage{xspace}  
\usepackage{color}
\usepackage{colortbl}

\usepackage{dcolumn}
\newboolean{pdflatex}
\setboolean{pdflatex}{true} 
\newboolean{uprightparticles}
\setboolean{uprightparticles}{false} 
\usepackage{amssymb}
\usepackage{amsfonts}
\usepackage{upgreek}
\usepackage{rotating}
\usepackage{multirow}

\hypersetup{ pdfborder=0 0 0, urlbordercolor=1 0 0 }

\usepackage{cite}

\renewcommand{\cite}[1]{\citep{#1}}

\def\lhcb {LHCb\xspace}
\def\ux85 {UX85\xspace}

\ifthenelse{\boolean{uprightparticles}}%
{

 \def\Pmu         {\ensuremath{\upmu}\xspace}

 \def\Ppi         {\ensuremath{\uppi}\xspace}

 \def\Ppsi        {\ensuremath{\uppsi}\xspace}

 \def\PDelta      {\ensuremath{\Delta}\xspace}                 
 \def\PXi      {\ensuremath{\Xi}\xspace}                 
 \def\PLambda      {\ensuremath{\Lambda}\xspace}                 
 \def\PSigma      {\ensuremath{\Sigma}\xspace}                 
 \def\POmega      {\ensuremath{\Omega}\xspace}                 
 \def\PUpsilon      {\ensuremath{\Upsilon}\xspace}                 
 
 \def\PB      {\ensuremath{\mathrm{B}}\xspace}                 
                  
 \def\PD      {\ensuremath{\mathrm{D}}\xspace}

 \def\PJ      {\ensuremath{\mathrm{J}}\xspace}                 
 \def\PK      {\ensuremath{\mathrm{K}}\xspace}

 \def\Pc      {\ensuremath{\mathrm{c}}\xspace}

 \def\Pi      {\ensuremath{\mathrm{i}}\xspace}

 \def\Ps      {\ensuremath{\mathrm{s}}\xspace}

}
{

 \def\Pmu         {\ensuremath{\mu}\xspace}

 \def\Ppi         {\ensuremath{\pi}\xspace}

 \def\Ppsi        {\ensuremath{\psi}\xspace}                 
                  
 \mathchardef\PDelta="7101
 \mathchardef\PXi="7104
 \mathchardef\PLambda="7103
 \mathchardef\PSigma="7106
 \mathchardef\POmega="710A
 \mathchardef\PUpsilon="7107
                  
 \def\PB      {\ensuremath{B}\xspace}                 
                  
 \def\PD      {\ensuremath{D}\xspace}

 \def\PJ      {\ensuremath{J}\xspace}                 
 \def\PK      {\ensuremath{K}\xspace}

 \def\Pc      {\ensuremath{c}\xspace}

 \def\Pi      {\ensuremath{i}\xspace}

 \def\Ps      {\ensuremath{s}\xspace}

}

\def\mumu       {\ensuremath{\Pmu^+\Pmu^-}\xspace}

\def\s     {\ensuremath{\Ps}\xspace}

\def\c     {\ensuremath{\Pc}\xspace}

\def\pion  {\ensuremath{\Ppi}\xspace}

\def\pip   {\ensuremath{\pion^+}\xspace}
\def\pim   {\ensuremath{\pion^-}\xspace}

\def\pipm  {\ensuremath{\pion^\pm}\xspace}

\def\kaon  {\ensuremath{\PK}\xspace}
  \def\Kbar  {\kern 0.2em\overline{\kern -0.2em \PK}{}\xspace}

\def\Kz    {\ensuremath{\kaon^0}\xspace}
\def\Kzb   {\ensuremath{\Kbar^0}\xspace}
\def\KzKzb {\ensuremath{\Kz \kern -0.16em \Kzb}\xspace}
\def\Kp    {\ensuremath{\kaon^+}\xspace}
\def\Km    {\ensuremath{\kaon^-}\xspace}

\def\KpKm  {\ensuremath{\Kp \kern -0.16em \Km}\xspace}

\def\Kstarz  {\ensuremath{\kaon^{*0}}\xspace}
\def\Kstarzb {\ensuremath{\Kbar^{*0}}\xspace}
\def\Kstar   {\ensuremath{\kaon^*}\xspace}

  \def\Dbar    {\kern 0.2em\overline{\kern -0.2em \PD}{}\xspace}
\def\D       {\ensuremath{\PD}\xspace}

\def\Dz      {\ensuremath{\D^0}\xspace}
\def\Dzb     {\ensuremath{\Dbar^0}\xspace}
\def\DzDzb   {\ensuremath{\Dz {\kern -0.16em \Dzb}}\xspace}
\def\Dp      {\ensuremath{\D^+}\xspace}
\def\Dm      {\ensuremath{\D^-}\xspace}

\def\DpDm    {\ensuremath{\Dp {\kern -0.16em \Dm}}\xspace}

\def\B       {\ensuremath{\PB}\xspace}
  \def\Bbar    {\kern 0.18em\overline{\kern -0.18em \PB}{}\xspace}

\def\Bz      {\ensuremath{\B^0}\xspace}

\def\Bd      {\ensuremath{\B^0}\xspace}
\def\Bs      {\ensuremath{\B^0_s}\xspace}

\def\jpsi     {\ensuremath{{\PJ\mskip -3mu/\mskip -2mu\Ppsi\mskip 2mu}}\xspace}

  \def\Y#1S{\ensuremath{\PUpsilon{(#1S)}}\xspace}

\def\BR         {{\ensuremath{\cal B}\xspace}}

\def\to                 {\ensuremath{\rightarrow}\xspace}

\def\CP                {\ensuremath{C\!P}\xspace}

\newcommand{\betas}{\ensuremath{\beta_{\s}}\xspace}

\def\AT#1     {\ensuremath{A_T^{#1}}\xspace}           

\def\C#1      {\ensuremath{\mathcal{C}_{#1}}\xspace}                       
\def\Cp#1     {\ensuremath{\mathcal{C}_{#1}^{'}}\xspace}                    
\def\Ceff#1   {\ensuremath{\mathcal{C}_{#1}^{\mathrm{(eff)}}}\xspace}        
\def\Cpeff#1  {\ensuremath{\mathcal{C}_{#1}^{'\mathrm{(eff)}}}\xspace}       
\def\Ope#1    {\ensuremath{\mathcal{O}_{#1}}\xspace}                       
\def\Opep#1   {\ensuremath{\mathcal{O}_{#1}^{'}}\xspace}                    

\newcommand{\tev}{\ensuremath{\mathrm{\,Te\kern -0.1em V}}\xspace}
\newcommand{\gev}{\ensuremath{\mathrm{\,Ge\kern -0.1em V}}\xspace}
\newcommand{\mev}{\ensuremath{\mathrm{\,Me\kern -0.1em V}}\xspace}
\newcommand{\kev}{\ensuremath{\mathrm{\,ke\kern -0.1em V}}\xspace}
\newcommand{\ev}{\ensuremath{\mathrm{\,e\kern -0.1em V}}\xspace}
\newcommand{\gevc}{\ensuremath{{\mathrm{\,Ge\kern -0.1em V\!/}c}}\xspace}
\newcommand{\mevc}{\ensuremath{{\mathrm{\,Me\kern -0.1em V\!/}c}}\xspace}
\newcommand{\gevcc}{\ensuremath{{\mathrm{\,Ge\kern -0.1em V\!/}c^2}}\xspace}
\newcommand{\gevgevcccc}{\ensuremath{{\mathrm{\,Ge\kern -0.1em V^2\!/}c^4}}\xspace}
\newcommand{\mevcc}{\ensuremath{{\mathrm{\,Me\kern -0.1em V\!/}c^2}}\xspace}

\def\invpb {\ensuremath{\mbox{\,pb}^{-1}}\xspace}

\newcommand{\chisq}{\ensuremath{\chi^2}\xspace}

\def\gsim{{~\raise.15em\hbox{$>$}\kern-.85em
          \lower.35em\hbox{$\sim$}~}\xspace}
\def\lsim{{~\raise.15em\hbox{$<$}\kern-.85em
          \lower.35em\hbox{$\sim$}~}\xspace}

\def\ptot       {\mbox{$p$}\xspace}
\def\pt         {\mbox{$p_T$}\xspace}

\def\tell1  {TELL1\xspace}
\def\ukl1   {UKL1\xspace}

\begin{document}

\begin{titlepage}

\belowpdfbookmark{Title page}{title}

\pagenumbering{roman}

\vspace*{-1.5cm}

\centerline{\large EUROPEAN ORGANIZATION FOR NUCLEAR RESEARCH (CERN)}

\vspace*{1.5cm}

\hspace*{-5mm}\begin{tabular*}{16cm}{lc@{\extracolsep{\fill}}r}

\vspace*{-12mm}\mbox{\!\!\!\epsfig{figure=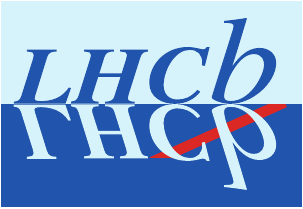,width=.12\textwidth}}&
& \\

&& LHCb-PAPER-2011-012\\
&& CERN-PH-EP-2011-183\\

&&  January 9, 2012 \\

\end{tabular*}

\vspace*{4cm}

\begin{center}

{\bf\huge\boldmath{First observation of the decay $\Bs \to \Kstarz \Kstarzb$}\\
}

\vspace*{2cm}

\normalsize {

The LHCb Collaboration\footnote{Authors are listed on the following pages.}


}

\end{center}

\vspace{\fill}

\centerline{\bf Abstract}

\vspace*{5mm}

\noindent The first observation of the decay $\kstarkstar$ is reported using
35\invpb of data collected by LHCb in proton-proton collisions at a
centre-of-mass energy of 7 TeV. A total of $49.8 \pm 7.5$ $B^0_s \rightarrow (K^+\pi^-)(K^-\pi^+)$ events are
{observed within $\pm 50 \mevcc$ of the \Bs mass and $746 \mevcc < m_{K\pi}< 1046 \mevcc$,
mostly coming from a resonant $\kstarkstar$ signal.}  The branching fraction and the
\CP-averaged \Kstarz longitudinal polarization fraction are measured to be
{$\BR\left( \kstarkstar\right) = (2.81 \pm 0.46 ({\rm stat.}) \pm 0.45 ({\rm
syst.}) \pm 0.34\, (f_s/f_d) )\times10^{-5}$} and $f_L = 0.31 \pm 0.12 ({\rm
stat.}) \pm 0.04 ({\rm syst.})$.

\vspace*{1.cm}


\vspace{\fill}
\centerline{(Submitted to Physics Letters B.)}
\vspace*{0.5cm}

\clearpage

\setcounter{page}{2}

\belowpdfbookmark{LHCb author list}{authors}


\noindent
R.~Aaij$^{23}$, 
C.~Abellan~Beteta$^{35,n}$, 
B.~Adeva$^{36}$, 
M.~Adinolfi$^{42}$, 
C.~Adrover$^{6}$, 
A.~Affolder$^{48}$, 
Z.~Ajaltouni$^{5}$, 
J.~Albrecht$^{37}$, 
F.~Alessio$^{37}$, 
M.~Alexander$^{47}$
% ...

\noindent
{\it \footnotesize
% ...
$ ^{5}$Clermont Universit\'{e}, Universit\'{e} Blaise Pascal, CNRS/IN2P3, LPC, Clermont-Ferrand, France\\
$ ^{6}$CPPM, Aix-Marseille Universit\'{e}, CNRS/IN2P3, Marseille, France\\
% ...
$ ^{23}$Nikhef National Institute for Subatomic Physics, Amsterdam, Netherlands\\
% ...
$ ^{35}$Universitat de Barcelona, Barcelona, Spain\\
$ ^{36}$Universidad de Santiago de Compostela, Santiago de Compostela, Spain\\
$ ^{37}$European Organization for Nuclear Research (CERN), Geneva, Switzerland\\
% ...
$ ^{42}$H.H. Wills Physics Laboratory, University of Bristol, Bristol, United Kingdom\\
% ...
$ ^{47}$School of Physics and Astronomy, University of Glasgow, Glasgow, United Kingdom\\
$ ^{48}$Oliver Lodge Laboratory, University of Liverpool, Liverpool, United Kingdom\\
% ...
$ ^{n}$LIFAELS, La Salle, Universitat Ramon Llull, Barcelona, Spain\\
% ...
}

\end{titlepage}

\cleardoublepage

\setcounter{page}{1}

\pagenumbering{arabic}


\pagestyle{plain} 
\setcounter{page}{1}
\pagenumbering{arabic}


\section{Introduction}
\label{sec:Introduction}

The decay $\kstarkstar$ is described in the Standard Model by loop (penguin)
diagrams that contain a $b \to s$ transition.  The partial width of the decay
arises from three helicity amplitudes that, assuming no additional contributions
from physics beyond the Standard Model, are determined by the chiral structure
of the quark operators.  Predictions obtained within the framework of QCD
factorization~\cite{beneke} are $\BR(\kstarkstar) = (9.1^{+11.3}_{-6.8}) \times
10^{-6}$ for the branching fraction and ${0.63}^{+0.42}_{-0.29}$ for the \Kstarz
longitudinal polarization fraction. Predictions  improve to
$({7.9}^{+4.3}_{-3.9}) \times 10^{-6}$ and ${0.72}^{+0.16}_{-0.21}$,
respectively, when experimental input is used from  $\B \to \Kstar\phi$
\cite{babar_kstphi,belle_kstphi}.  The possibility to use $\kstarkstar$ for
precision \CP-violation studies to determine the phases \betas and $\gamma$ of
the CKM matrix \cite{pdg} has been emphasized by several authors
\cite{matias,ciuchini,fleischer,fleischer99}.  The U-spin related channel, $\Bd
\to \Kstarz \Kstarzb$, a $b \to d$ transition, has been observed by
BaBar~\cite{bbar}, reporting a branching fraction of $({1.28}^{+0.35}_{-0.30}
\pm 0.11) \times 10^{-6}$ and $f_L = {0.80}^{+0.10}_{-0.12} \pm 0.06$ with a
signal yield of ${33.5}^{+9.1}_{-8.1}$ events. An upper limit  for the
$\kstarkstar$  branching fraction of $1.68 \times {10}^{-3}$ with $90\%$
confidence level was reported by the SLD experiment \cite{sld}. 

We present in this Letter the first observation of the $\kstarkstar$ decay using
$pp$ collisions at  $\sqrt{s} = 7$ TeV at the LHC. The data were collected
during 2010 and corresponds to 35 \invpb of integrated luminosity.  LHCb has
excellent capabilities to trigger and reconstruct beauty and charm hadrons, and
covers the pseudorapidity region $2<\eta<5$. The tracking system consists of a
21 station, 1-metre long array of silicon strip detectors placed within 8 mm of
the LHC beams. This is followed by a four layer silicon strip detector upstream
of a 4 Tm dipole magnet. Downstream of the magnet are three tracking stations,
each composed of a four-layer silicon strip detector in the high occupancy
region near the beam pipe, and an eight layer straw tube drift chamber composed
of 5 mm diameter straws outside this high occupancy region. Overall, the
tracking system provides an impact parameter (IP) \footnote{The impact parameter
is the distance of closest approach between a particle's trajectory and its
assumed production point.} resolution of 16 $\mu$m + 30 $\mu$m/$p_{\rm{T}}
(\gevc)$, and a momentum resolution $\sigma_p/p$ below 8 per mille up to 100
\gevc. Two ring imaging Cherenkov detectors, one upstream of the magnet, and a
second just downstream of the tracking stations, together provide a typical kaon
identification efficiency of 90\%. The  pion fake rate, over the momentum range
from $3-100$~\gevc, is between 5 and 10 percent. Further downstream is a
Preshower/Scintillating Pad Detector, an electromagnetic calorimeter, and a
hadronic calorimeter. The LHCb spectrometer also features a large, five station
muon system used for triggering on and identifying muons. A more detailed
description of the LHCb detector can be found in~\cite{jinst}.

To reduce the data rate from the LHC crossing rate to about 2 kHz for permanent
storage, LHCb uses a two-level trigger system. The first level of the trigger,
implemented in hardware,
searches for either a large transverse energy ($E_{\rm{T}}$) cluster  in the
calorimeters ($E_{\rm{T}}>3.6$ GeV is a representative value during the 2010 run), or a
single high transverse momentum ($p_{\rm{T}}$) muon or di-muon pair in the muon
stations. 

Events passing the hardware trigger are read out and sent to a large computing
farm, where they are analyzed using a software-based trigger~\cite{HLT2}.  The
first stage of the software trigger relies on the selection of a single track
with IP larger than 125 $\mu$m, $p_{\rm{T}}> 1.8$ \gevc, $p > 12.5$ \gevc, along
with other track quality requirements. Events are subsequently analyzed by a
second software stage, where the event is searched for 2, 3, or 4-particle
vertices that are consistent with originating from $b$-hadron decays.  The
impact parameter $\chi^2$ of the selected tracks (IP$\chi^2$), defined as the
difference between the $\chi^2$ of the primary vertex (PV) built with and
without the considered track, is required to be greater than 16 with respect to
any PV. The tracks are also required to have $p>5$~\gevc and $p_{\rm{T}} >
0.5$~\gevc. The \Bs decay vertex must have at least one track with
$p_{\rm{T}}>1.5$~\gevc, a scalar $p_{\rm{T}}$ sum of at least 4 \gevc, and a
corrected mass\footnote{The corrected mass is related to the invariant mass $m$,
as $m_{corr} = \sqrt{m^2 + |p_{T miss}|^2} + |p_{T miss}|$ , where $p_{T miss}$
is the missing momentum transverse to the \Bs direction.} between 4 and 7
\gevcc. Additional track and vertex quality cuts are also applied. 

Events with large occupancy are slow to reconstruct and were suppressed by
applying global event cuts to hadronically triggered decays. These included
limits on the number of hits in the tracking detectors and scintillating pad
detector.

\section{Selection procedure and signal yield}
\label{sec:selyield}

To search for the decay process $\Bs \to \Kstarz(\Kp\pim) \Kstarzb(\Km\pip)$ we
applied a number of offline selection criteria.  When a four-track secondary
vertex is found, the reconstructed momentum of the \Bs candidate is used to
calculate the smallest impact parameter with respect to all primary vertices in
the event. Tracks are required to have $p_{\rm{T}} > 500$~\mevc, and a large
impact parameter (IP$\chi^2>9$) with respect to the PV.  The difference in the
natural logarithm of the likelihoods of the kaon and pion hypotheses must be
greater than 2 for $K^{+}$ and $K^{-}$ candidates, and less than 0 for $\pi^{+}$
and $\pi^{-}$ candidates. In addition, the $K^+ \pi^-$
combinations\footnote{This expression refers hereafter to both charge
combinations: $K^+ \pi^-$ and $K^- \pi^+$.} must form an acceptable quality
common vertex ($\chi^{2}$/ndf$~<9$), where ndf is the number of degrees of
freedom in the vertex fit) and must have an invariant mass within $\pm150$
\mevcc of the nominal \Kstarz mass (this is around $\pm$3 times its physical
width~\cite{pdg}). The \Kstarz and \Kstarzb candidates must have $p_{\rm{T}} >
900$~\mevc and the distance of closest approach between their trajectories must
be less than 0.3 mm.  The secondary vertex must be well fitted
($\chi^{2}/$ndf$<5$).  Finally, the \Bs candidate momentum is required to point
to the PV.

\begin{figure}[htbp]
  \begin{center}
  \ifthenelse{\boolean{pdflatex}}{
    \includegraphics*[width=\columnwidth]{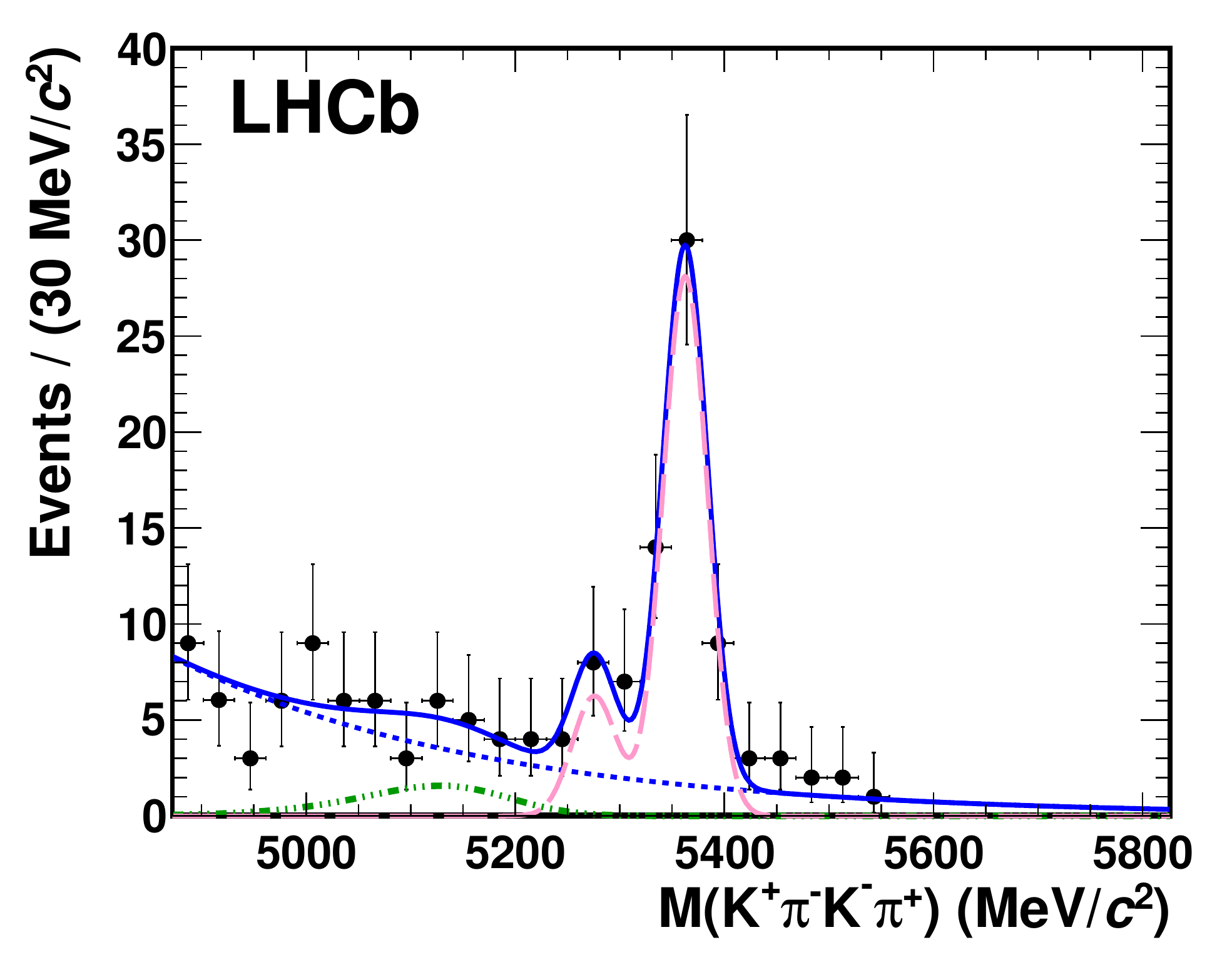}
   }{
    \includegraphics*[width=\columnwidth]{Fig1.eps}
   }
  \end{center}
\caption{Fit to the $\Kp\pim\Km\pip$ mass distribution of selected candidates.
The fit model (dashed pink curve) includes a signal component that has two
Gaussian components corresponding to the \Bs and \Bd decays. The background is
described as an exponential component (dotted blue) plus the parametrization
indicated in the text (dash-dotted green).}
\label{figBsmass}
\end{figure}

To improve the signal significance, a multivariate analysis is used that takes
into account the properties of the $\Bs \to \Kstarz(\Kp\pim) \Kstarzb(\Km\pip)$
signal, as well as those of the background. It is based on a geometrical
likelihood (GL)~\cite{Karlen,DiegoThesis} that uses the following set of
variables 
as input:

\begin{itemize}

\item \Bs candidate impact parameter with respect to the closest  primary vertex.

\item Decay time of the \Bs candidate.

\item Minimum impact parameter \chisq of the four tracks with respect to all
primary vertices in the event.

\item Distance of closest approach between the two \Kstarz trajectories
reconstructed from the pion and kaon tracks. 

\item $p_{\rm{T}}$ of the \Bs candidate.

\end{itemize}

For a given input sample, the above distributions are converted into a set of
uncorrelated, Gaussian-distributed variables.  Two vectors are defined for each
event indicating its distance to the signal $\{S_i\}$ and to the background
$\{B_i\}$ hypotheses by means of $\chi^2_S = \sum S_i^2$  and   $\chi^2_B = \sum
B_i^2$, {where the index $i$ runs over the five discriminating variables indicated above.}
The quantity $\Delta\chi^2 = \chi^2_S-\chi^2_B$ is found to be a good
discriminant between the two hypotheses and is used to construct the GL function
in such a way that it is uniformly distributed in the range $[0,1]$ for signal
events and tends to have low values for the background. {The signal input is generated 
by EvtGen \cite{evtgen} and Pythia 6.4 \cite{pythia} for the kinematic spectra, and the full detector simulation is based on GEANT4 \cite{geant4}.} 

The GL selection requirement was determined by maximizing the signal
significance.  The GL was trained using a fully reconstructed $\kstarkstar$
simulation sample for the signal, and a selected background sample from the
first 2 \invpb of data, which is not used in the analysis.  The requirement
GL$>$0.24, together with the above selection criteria, resulted in the mass
spectrum in Fig.~\ref{figBsmass} for the selected $\Kp\pim\Km\pip$ candidates.
It is observed that the events with masses below the signal region have on
average slightly higher GL values than those with masses above. This indicates
the presence of a background from partially reconstructed \B decays. 

To describe the data, we have used a fit including two Gaussian probability
density functions (PDFs) centered at the \Bd and \Bs masses respectively, a
decreasing exponential and the following parametrization for partially
reconstructed \mbox{\B-decays}
  \begin{equation}
    \begin{small}
    \label{eq:physbkg}
    A M^{\prime}
    \left(1-\frac{M^{\prime 2}}{M_p^2}\right)
    \Theta(M_p-M^{\prime})
    e^{-k_p \cdot M^{\prime}}
    \otimes
    G(M-M^{\prime};\sigma_p),
    \end{small}
  \end{equation}
where $\Theta$ is the Heaviside-step function, $\otimes$ represents the
convolution, $M^{\prime}$ is the variable over which the convolution integral is
calculated, $G(M - M^{\prime}; \sigma_p)$ is a Gaussian PDF with standard
deviation $\sigma_p$ and $M_p$ and $k_p$ are free parameters. The fit results
are given in Table \ref{tab:massfitresults}.

The measured signal yield in a window of $\pm 50 \mevcc$ around the \Bs mass is
$N_{\Kp\pim\Km\pip} = 49.8 \pm 7.5 ({\rm stat.})$. The width of the \Bs peak is
in good agreement with the \lhcb resolution measured in decays with similar
kinematics such as $\Bs \to \jpsi\phi$. The significance of the \Bs signal was
determined to be $10.9~\sigma$ by comparing the log of the likelihood between
the models with and without signal. When doing this test, the mass and width of
the \Bs and \Bd mesons were fixed to those obtained from independent LHCb
measurements of $\Bs \to \jpsi\phi$ and $\jpsikstar$, respectively. The peak at
the \Bd mass, though not significant, is compatible with the $\Bdkstarkstar$
branching fraction measured by BaBar~\cite{bbar}.

\begin{table}
\caption{Fitted values of the model parameters for the mass spectrum, as
described in the text. $N_s$ and $N_d$ are the number of events for the \Bs and
\Bd signals, $\mu_s$ is the fitted mass value for the \Bs signal and $\sigma$ is
the Gaussian width. The mass difference between \Bs and \Bd was fixed to its
nominal value~\cite{pdg}. $N_{b}$ is the number of background events in the full
mass range (4900-5800 \mevcc), and $c_{b}$ is the exponential parameter in the
fit. $M_p$, $\sigma_p$ and $k_p$ are the parameters of Eq.~(\ref{eq:physbkg}).
Finally, $f_{p}$ is the fraction of the background associated with
Eq.~(\ref{eq:physbkg}). }
\label{tab:massfitresults}

\begin{center}
\newcolumntype{+}{D{/}{\ \pm\ }{-1}}
\begin{tabular}{|l|+|}
\hline
Parameter & \multicolumn{1}{c|}{Value}\\
\hline
$N_s$ & 50.1/7.5 \\
$N_d$ & 11.2/4.3 \\
$\mu_s$ {\small(\mevcc)} & 5362.5/4.8 \\
$\sigma$ {\small(\mevcc)} & 21.2/3.3  \\
$N_{b}$ & 90/10 \\
$c_{b}$ {\small($10^{-3}(\mevcc)^{-1}$)} & -3.37/0.55  \\
$k_p$ {\small($10^{-2}(\mevcc)^{-1}$)}& 5.5/5.3  \\
$f_{p}$ & \multicolumn{1}{c|}{${}_{\ \ }0.06_{\ -\ 0.05}^{\ +\ 0.24}$}  \\
$M_p$ {\small(\mevcc)} & 5170/170  \\
$\sigma_p$ {\small(\mevcc)} & 37/23  \\
\hline
\end{tabular}
\end{center}
\label{tab:massfit}
\end{table} 

As the \Kstarz meson is light compared to the \Bs meson, the invariant masses of
the three-body systems $\Kp\Km\pipm$ and $\Kp\pim\pipm$ are rather high, above
those of the charmed hadrons. This kinematically excludes the possibility of
contamination from $b \to c$ decays with very short charm flight distance, in
particular $\Bs \to D_s^- \pi^+$.

\begin{figure}[htbp]
  \begin{center}
  \ifthenelse{\boolean{pdflatex}}{
     \includegraphics*[width=\columnwidth]{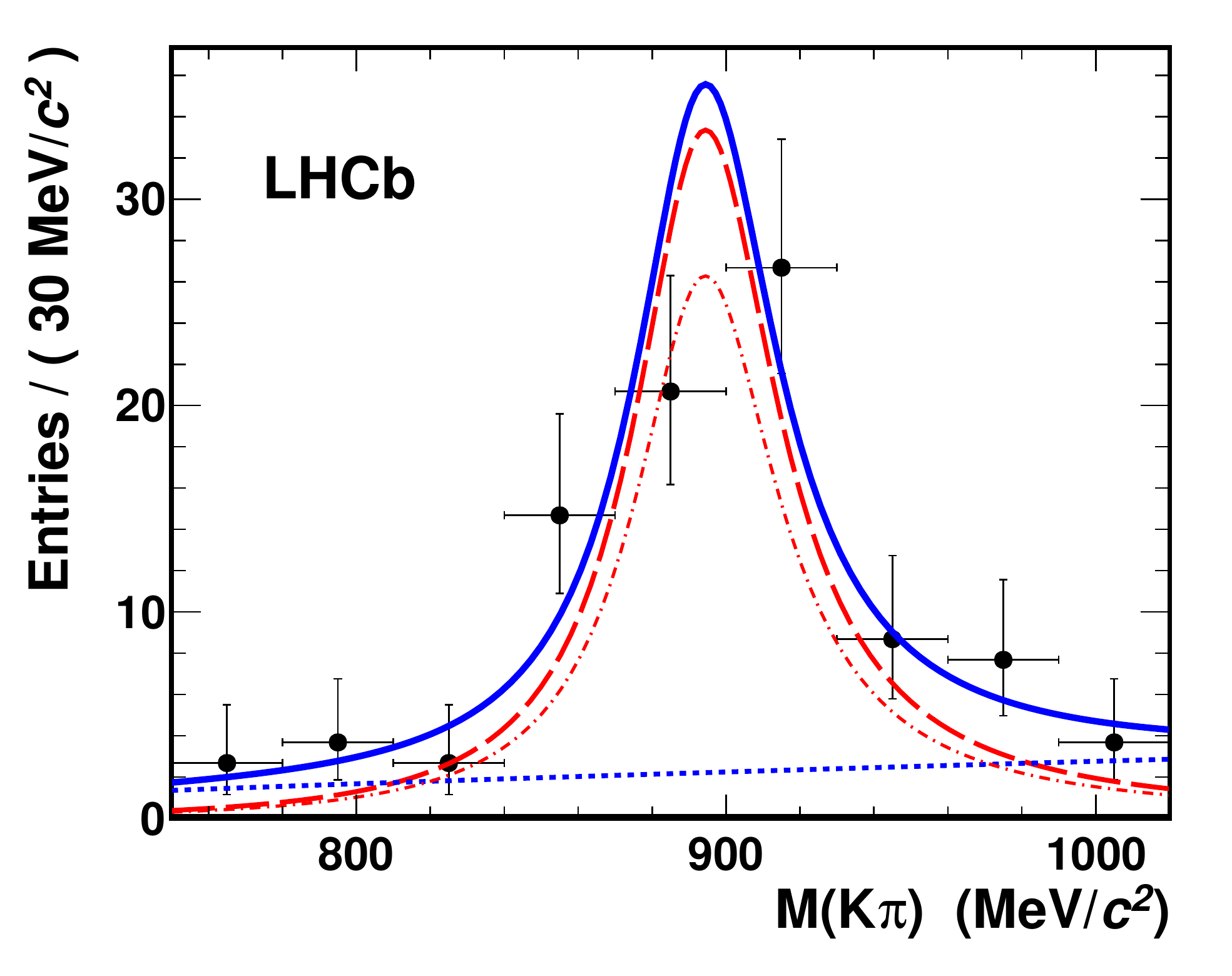}
   }{
     \includegraphics*[width=\columnwidth]{Fig2.eps}
   }
  \end{center}

\caption{Background subtracted $K^+\pi^-$ and $K^-\pi^+$ combinations for
selected candidates within a $\pm$50 \mevcc window of the \Bs mass. The solid
blue line shows the projection of the 2D fit model described in the text,
indicating the \Kstarz\Kstarzb yield (dashed-dotted red  line) and a nonresonant
component (blue dotted line), assumed to be a linear function times the two-body
phase space.  The dashed red line indicates the overall $\Bs\rightarrow\Kstarz
X$ contribution.}
\label{m1m2fit}
\end{figure}

After subtraction of the {non-\Bs component}, the $\Kp\pim$ mass combinations  were studied, within a
$\pm$50 \mevcc  window of the \Bs signal, by means of a maximum likelihood fit
in the $(m_{\Kp\pim},m_{\Km\pip})$ plane. Three components are included in the
fit, namely a double Breit-Wigner distribution describing $\kstarkstar$
production, a symmetrized product of a Breit-Wigner and a nonresonant linear
model adjusted for phase space in the $\Kp\pim$ mass, and a double nonresonant
component.  The fit result, as shown in Fig.~\ref{m1m2fit}, gives (62$\pm$18)\%
$\Kstarz\Kstarzb$ production.  The remainder is the symmetrized Breit-Wigner/
nonresonant model.

The shape of the background mass distribution was extracted from a fit to the
$\Kp\pim$ mass spectrum observed in two 400 \mevcc wide sidebands below and
above the \Bs mass.  The number of background events to be subtracted was
determined from the results in Table~\ref{tab:massfitresults}. The sizable
\Kstarz contribution present in this background was taken into account.

A model for $\Bs \to \Kstarz\Kstarzb$(1430), representing a broad scalar state
interfering with \kstarkstar was also studied in the available $K^+\pi^-$  mass
range of $\pm 150\mevcc$ around the \Kstarz mass.  The small number of events
made it impossible to measure precisely the size of such a contribution for all
values of the interfering phase.  However, for values of the phase away from
$\pi/2$ and $3\pi/2$ it was determined to be below 12\%.  Further study of this
issue requires a larger data sample.

\section{Selection of the control channel}

The branching fraction measurement of $\kstarkstar$ is based upon the use of a
normalization channel with a well measured branching fraction, and knowledge of
the selection and trigger efficiencies for both the signal and normalization
channels.  We chose $\jpsikstar$, with $\jpsi \to \mu^+ \mu^-$, for this
purpose. This decay has a similar topology to the signal, allowing the selection
cuts to be harmonized, and it is copiously produced in the LHCb acceptance. The
presence of two muons in the final state means that $\jpsikstar$ tends to be
triggered by a muon rather than a hadron, leading to a higher efficiency than
for $\kstarkstar$. The differences in the trigger can be mitigated by only
considering $\jpsikstar$ candidates where the trigger decision was not allowed
to be based on muon triggers that use tracks from the decay itself.

\begin{figure}
\begin{center}
\includegraphics[width=\columnwidth]{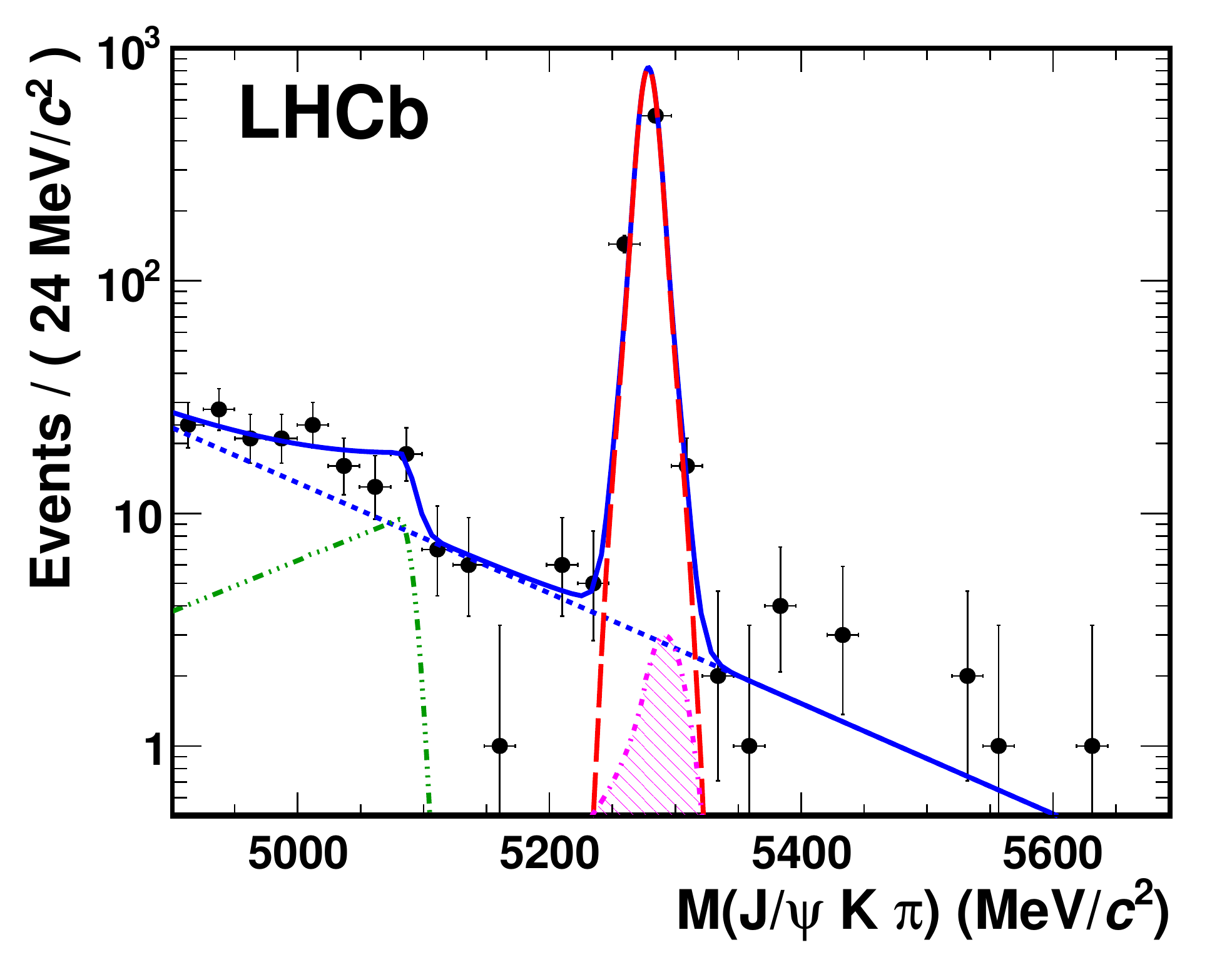}
\caption{Fit to the mass distribution of selected $\jpsikstar$ events. The
dashed red curve is the Gaussian component for the \B signal. The green
dashed-dotted line accounts for partially reconstructed \BJpsiX (see
Eq.~\ref{eq:BJpsiX}). The pink hatched region accounts for a possible $\Bs \to \jpsi \phi$
contamination, parametrized as a sum of two Crystal-Ball functions
\cite{cb}. The combinatorial background is parametrized as an exponential and
indicated as a blue dotted line.}
\label{fig:jpsikstarnew}
\end{center}
\end{figure}

The offline selection criteria for $\jpsikstar$ were designed to mimic those of
\kstarkstar. In particular, all cuts related to the \Bs vertex definition were
kept the same.  We also used the same GL as for the signal.

The overall detection efficiency was factorized as
$\epsilon^{sel}\epsilon^{trig}$. The first factor $\epsilon^{sel}$ is the
probability of the generated tracks being accepted in the \lhcb angular
coverage, reconstructed, and selected. The second factor $\epsilon^{trig}$
defines the efficiency of the trigger on the selected events.  Both are
indicated in Table~\ref{tab:BR_numbers}, as calculated from Monte Carlo
simulation, along with the number of selected events.  Note that our measurement
depends only on the ratios of efficiencies between signal and control channels.

\begin{table*}[ht]
  \begin{center}
    \caption{Selection and trigger efficiencies obtained from simulation.  The
observed yield found for the signal and control channels in the full mass range
are also indicated. The efficiency errors are statistical, derived from the size
of the simulated samples.}
    \newcolumntype{+}{D{/}{\ \pm\ }{5}}
    \begin{tabular}{|c|+|+|+|}
      \cline{2-4}
      \multicolumn{1}{c|}{} & \multicolumn{1}{c|}{$\epsilon^{sel}$ $(\%)$} & \multicolumn{1}{c|}{$\epsilon^{trig}$ $(\%)$} &  
\multicolumn{1}{c|}{\begin{minipage}{0.9in} \begin{center}
\medskip 
Yield
\medskip  \end{center} \end{minipage}} \\
      \hline
      \begin{minipage}{1.3in}  \begin{center} \smallskip $\kstarkstar$
  \smallskip \end{center} \end{minipage} & 0.370/0.005 &  37.12/0.39 & 42.5/6.7 \\
      \begin{minipage}{1.3in}  \begin{center} \smallskip $\jpsikstar$ \smallskip
\end{center} \end{minipage} & 0.547/0.007 &  31.16/0.63 & 657/27 \\
      \hline
      \begin{minipage}{1.3in}  \begin{center} \smallskip ratio \smallskip
 \end{center} \end{minipage} & 0.678/0.013 & 1.191/0.027 & 0.065/0.011\\
      \hline
    \end{tabular}
    \label{tab:BR_numbers}
  \end{center}
\end{table*}

The event yield for the selected data was determined from a fit to the
$\jpsi\Kp\pim$ invariant mass spectrum as shown in Fig.~\ref{fig:jpsikstarnew}.
In this fit, a constrained \jpsi mass was used in order to improve the \Bd mass
resolution and therefore background rejection.  
{A component for the particular background source $\Bs \to \jpsi \phi$, with $\phi \to K^+K^-$,
was included in the fit, with a parametrization defined from simulation, yielding the result 8$\pm$8 events.
The complete suppression of this background was subsequently confirmed using }
the Armenteros-Podolanski \cite{armenteros} plot
for the \Kstarz kinematics. 
The fit model also includes a Gaussian signal for the \Bd meson, and a combinatorial
background component parameterized with an exponential function and an
additional component to account for partially reconstructed \BJpsiX~\cite{PK}.
This partially reconstructed component can be described as
\begin{equation}
  \begin{tiny}
    \rho(M, \overline{M}, \mu,\kappa) \propto \begin{cases}
      e^{-\frac{1}{2}(\frac{M-\overline{M}}{\kappa})^2} & \text{if } M > \mu ; \\
      e^{-\frac{1}{2}(\frac{\mu-\overline{M}}{\kappa})^2 +
        \frac{(\overline{M}-\mu)(M-\mu)}{\kappa^{2}}} & \text{if } M \leq \mu. \\
    \end{cases}
    \label{eq:BJpsiX}
  \end{tiny}
\end{equation}
where the parameters $\mu$, $\kappa$ and $\overline{M}$ are allowed to float.
{The fitted signal according to this model is indicated in the third column of 
Table~\ref{tab:BR_numbers}.}

A small fraction of the selected sample contains two alternative candidates for
the reconstructed event, which share three of the particles but differ in the
fourth one. Those events, which amount to 3.8 \%  (3.7\%) in the signal
(control) channels, were retained for the determination of the branching
fraction.

\section{Analysis of \Kstarz polarization}
\label{sec:Kpolarization}

The four-particle $K^+\pi^-K^-\pi^+$ angular distribution describing the decay
of \Bs into two vector mesons ($\Kstarz \to K^+\pi^-$ and $\Kstarzb \to
K^-\pi^+$) is determined by three transversity amplitudes $\mathcal{A}_L$,
$\mathcal{A}_{\parallel}$ and $\mathcal{A}_{\perp}$.  The relative fraction of
these can be determined from the distribution of the decay products in three
angles $\theta_1$, $\theta_2$ and $\varphi$. Here $\theta_1$ ($\theta_2$) is the
$K^+$ ($K^-$) emission angle with respect to the direction opposite to the \Bs
meson momentum in the \Kstarz (\Kstarzb) rest frame, and $\varphi$ is the angle between
the normals to the \Kstarz and \Kstarzb decay planes in the \Bs rest frame
\cite{matias}.  We will refer generically to the $\theta$ angle from now on,
unless differences between $\theta_1$ and $\theta_2$ become relevant for the
discussion. In a time-integrated and flavor-averaged analysis, and assuming the
\Bs mixing phase $\beta_s \approx 0$ as in the Standard Model, the angular
distribution is given by~\cite{matias,dunietz}

\begin{multline}
I(\theta_1,\theta_2,\varphi) = \frac{{\rm d}^3\Gamma}{{\rm d}\cos\theta_1\, {\rm
d} \cos\theta_2\, {\rm d}\varphi}  = \\ \left( \frac{1}{\Gamma_L} \right.
|\mathcal{A}_L|^2 \cos^2\theta_1 \cos^2\theta_2 \\ + \frac{1}{\Gamma_L}
|\mathcal{A}_\parallel|^2 \frac{1}{2} \sin^2\theta_1 \sin^2\theta_2
\cos^2\varphi \\ + \frac{1}{\Gamma_H} |\mathcal{A}_\perp|^2 \frac{1}{2}
\sin^2\theta_1 \sin^2\theta_2 \sin^2\varphi \\ + \frac{1}{\Gamma_L}
|\mathcal{A}_L| \left.|\mathcal{A}_\parallel| \cos\delta_\parallel
\frac{1}{2\sqrt{2}} \sin 2\theta_1 \sin 2\theta_2 \cos\varphi \right)
\label{eq:angular}
\end{multline}

\noindent 
We denote the polarization fractions by
\begin{equation}
  \begin{small}
    f_k = \frac{|\mathcal{A}_k|^2}{|\mathcal{A}_L|^2+|\mathcal{A}_\parallel|^2+
      |\mathcal{A}_\perp|^2} \ \ \ ,\ \ \  k=L, \parallel, \perp.
  \end{small}
\label{eq:amplitudes}
\end{equation}

\noindent and consequently $ f_L+f_{\parallel}+f_{\perp}=1 $.  No \CP violation
in the mixing or in the decay has been considered.  The interference terms
related to the $\mathcal{A}_{\perp}$ amplitude, both proportional to
sin$\phi_s$, have been neglected.  $\Gamma_{L,H}$ are the total widths of the
low and high mass eigenstates of the \Bs meson, respectively, and
$\delta_\parallel$ is the phase difference between $\mathcal{A}_L$ and
$\mathcal{A}_{\parallel}$. The total decay width is defined as $\Gamma =
(\Gamma_L + \Gamma_H)/2$ and $\Delta\Gamma=\Gamma_L-\Gamma_H$.  Note that as a
consequence of time integration the relative normalization acquired by the
\CP-even and \CP-odd terms is different. The values $\Delta\Gamma =
(0.062^{+0.034}_{-0.037}) \times 10^{12}\ \rm{s}^{-1}$ and $\Gamma =
(0.679^{+0.012}_{-0.011}) \times 10^{12}\ \rm{s}^{-1}$~\cite{pdg} were used. 

The detector acceptance is compatible with being constant in $\varphi$.  In
contrast, it has a significant dependence on the \Kstarz polarization angle
$\theta$.  The two-dimensional angular acceptance function
$\epsilon$($\cos\theta_1$,$\cos\theta_2$) was studied with a full detector
simulation. It drops to nearly zero asymmetrically as $\cos \theta_{1,2}$
becomes close to $\pm 1$, as a consequence of the minimum \ptot and \pt of the
tracks imposed  by the  reconstruction.

The Monte Carlo simulation of the \Kstarz acceptance was extensively
cross-checked using the \jpsikstar control channel, taking advantage of the fact
that the \Kstarz polarization in this channel was measured at the \B-factory
experiments \cite{jpsikstar_pol,swave-bbar}. The function
$\epsilon$($\cos\theta_1$,$\cos \theta_2$) has been projected onto the \Kstarz
and \Kstarzb axes separately, showing no appreciable difference, and a small
average correlation, given the size of the simulated sample. We have then used
the one-dimensional acceptance $\epsilon_{\theta}(\cos\theta)$  as the basis of
our analysis, and determined it in five bins of $\cos\theta$.  Since the
longitudinal polarization fraction for the \jpsikstar channel is well measured,
a comparison between data and simulation is possible.  Agreement was found
including variations of the angular distribution with longitudinal and
transverse \Kstarz momentum.  In the region $\cos\theta > 0.6$ these variations
were four times larger than for lower values of  $\cos\theta$.

The background $\cos \theta$ distribution was studied in two 200\mevcc
sidebands, defined below and above the \Bs signal region.  Like the signal, it
showed a dip close to $\cos \theta = +1$ and it was parameterized  as
$\epsilon_{\theta} \cdot (1+\beta \cos\theta)$. A one parameter fit for $\beta$
gives the result $\beta=-0.18\pm0.13$. 

An unbinned maximum likelihood fit was then performed to the data in a $\pm 50$
\mevcc window around the \Bs mass, in the region $\cos\theta < 0.6 $, according
to the PDF
\begin{eqnarray}
\lefteqn{F(\theta_1,\theta_2,\varphi) = (1-\alpha) \epsilon_{\theta}(\theta_1)  \epsilon_{\theta}(\theta_2) I(\theta_1,\theta_2,\varphi)} \nonumber\\
& & + \alpha    (1+\beta \cos\theta_1) (1+\beta\cos\theta_2)  \epsilon_{\theta}(\theta_1)  \epsilon_{\theta}(\theta_2). \nonumber \\
& & 
\label{eq:angularPDF}
\end{eqnarray}
The background fraction $\alpha$ was determined from the fit to the \Bs mass
spectrum described in Sec. \ref{sec:selyield}. Only three parameters were
allowed to vary in the fit, namely $f_L$, $f_{\parallel}$  and the phase
difference $\delta_{\parallel}$.

\begin{figure}
\centering
\includegraphics[width=\columnwidth]{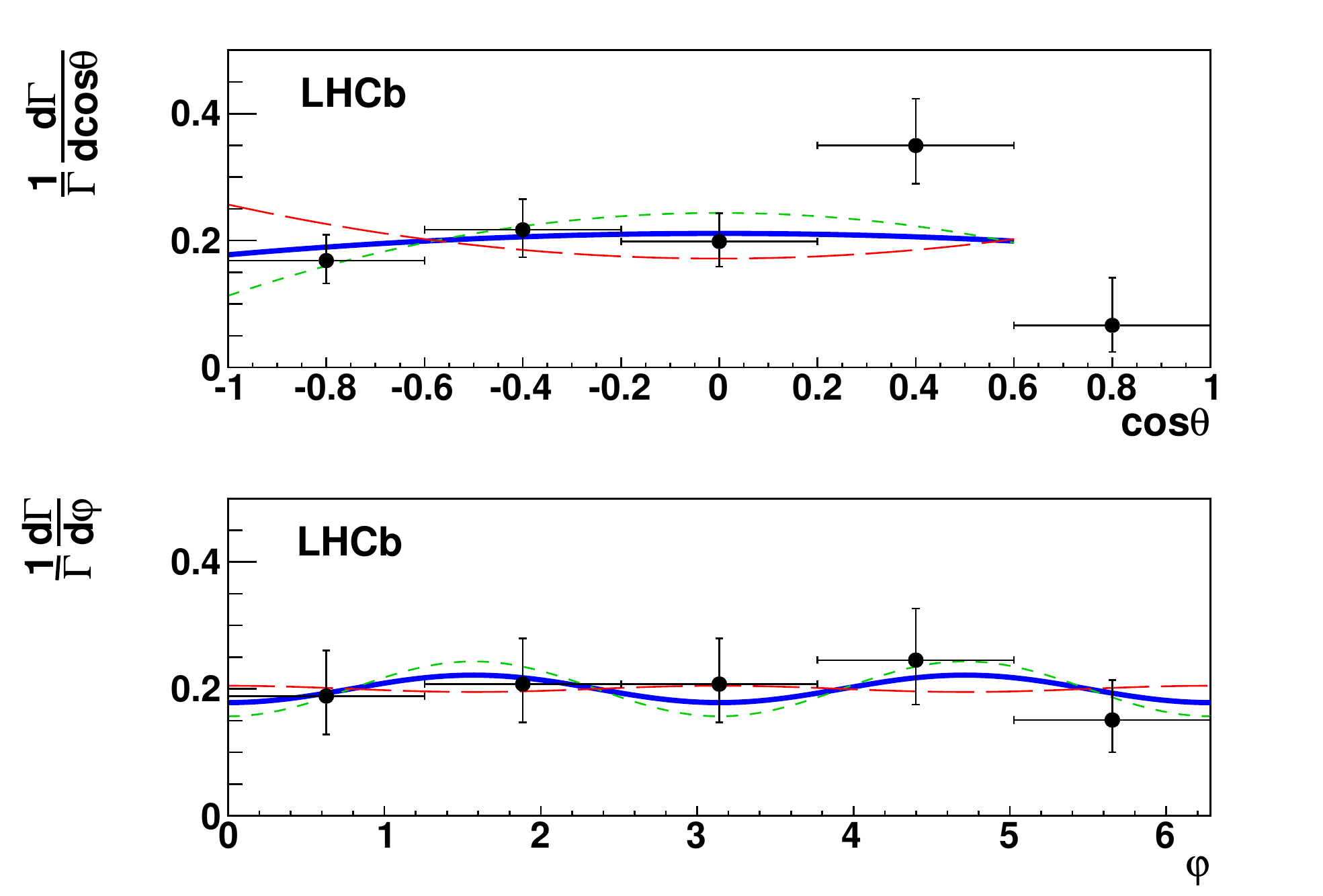}
\caption{$\cos\theta$ (above) and $\varphi$ (below) acceptance corrected
distributions for events in the narrow window around the \Bs mass. The blue line
is the projection of the fit model given by Eq.~\ref{eq:angular} for the
measured values of the parameters $f_L$, $f_{\parallel}$ and
$\delta_{\parallel}$. The dotted lines indicate $\pm 1\sigma$ variation of the
$f_L$ central value.}
\label{fig_final}
\end{figure}

One-dimensional projections of the fit results are shown in
Fig.~\ref{fig_final}.  The consistency of the measurement in various regions of
the \Kstarz phase space, and of the impact parameter of the daughter particles,
was checked. The experimental systematic error on $f_L$ was estimated from the
variation of the measurements amongst those regions to be $\pm$0.03.

The acceptance for \kstarkstar is not uniform as a function of proper decay time
due to the cuts made on the IP of the kaons and pions, and a small correction to
the polarization fractions, of order 3\%, was applied in order to take into
account this effect.  It was calculated from the variation in the measured
polarization amplitudes induced by including a parametrization of the time
acceptance in Eq.~\ref{eq:angularPDF}.  Note the different correction sign for
each polarization fraction, as a consequence of the assumption $\Delta\Gamma \ne
0$. 

The sensitivity of the $f_L$ measurement with respect to small variations of the
$\cos\theta$ distribution has been tested. These variations could be attributed
to experimental errors not accounted for in the simulation or to interference
with other partial waves in the $K\pi$ system.  A high statistics study using
\jpsikstar muon triggers revealed a small systematic difference between data and
simulation in $\epsilon_{\theta}(\cos\theta)$  as $\cos\theta$ approaches +1,
which was taken into account as a correction in our analysis.  When this
correction in varied by  $\pm 100$\%, $f_L$ varies by $\pm$0.02 which we
consider as an additional source of systematic error.  The total systematic on
$f_L$ is thus $\pm$0.04.

We finally measure the  \Kstarz  longitudinal polarization fraction $f_L =
0.31\pm 0.12 ({\rm stat.}) \pm 0.04 ({\rm syst.})$, as well as the transverse
components $f_{\parallel}$ and $f_{\perp}$.  In the small sample available, the
\CP-odd component $f_{\perp}$ appears to be sizable $f_{\perp} = 0.38 \pm 0.11
({\rm stat}.)\pm 0.04 ({\rm syst}.)$.  A significant measurement of
$\delta_{\parallel}$ could not be achieved ($\delta_{\parallel}$ = 1.47 $\pm$
1.85).

As seen in Eq.~(\ref{eq:angular}), due to a nonzero  $\Delta\Gamma$ time
integration changes the relative proportion between the various terms of the
angular distribution, with respect to their values at $t=0$.  If we call $
f_k^0$  the polarization fractions we would have measured under the assumption
$\Delta\Gamma = 0$, it can be derived from Eq. \ref{eq:angular} that our
measured values are
\begin{equation}
f_k = f_k^0 \left(1 + \eta_k \frac{\Delta\Gamma}{2\Gamma}\right)
\end{equation}
with CP eigenvalue $\eta_k = +1,+1,-1$ for $k=L,\parallel, \perp$. Given the
current knowledge of $\Delta\Gamma$/$\Gamma$ \cite{pdg}, the magnitude of the
correction to $f_k$ amounts to 4.6\%, and the associated systematic error
related to $\Delta\Gamma$ error is 2.6\%, which we have neglected in comparison
to other sources.

\section{Determination of the branching fraction}
\label{sec:branchingF}

The results of the previous sections can be brought together to provide a
determination of the branching fraction of the $\kstarkstar$ decay based upon
the use of the normalization channel $\jpsikstar$ through the expression
\begin{eqnarray}
\lefteqn{\BR\left( \kstarkstar\right)  = \lambda_{f_L} \times 
\frac{\epsilon^{sel}_{\jpsikstar}}{\epsilon^{sel}_{\kstarkstar}}} \nonumber \\
& & \times \frac{\epsilon^{trig}_{\jpsikstar}}{\epsilon^{trig}_{\kstarkstar}}  \times \frac{N_{\kstarkstar}}{N_{\jpsikstar}} \nonumber\\
& & \times \BR_{vis}{(\jpsikstar)} \times \frac{f_{d}}{f_s} \times \frac{9}{4},
\label{eqn:BR_BR_JPsiKstar}
\end{eqnarray}
where $\BR_{vis}{(\jpsikstar)}$, the visible branching ratio, is the product
$\BR(\jpsikstar) \times \BR(\jpsi \to \mumu) \times \BR(\Kstarz \to \Kp \pim)$.
The numerical value of $\BR(\jpsikstar) = (1.33 \pm 0.06) \times 10^{-3}$ is
taken from the world average in~\cite{pdg}, $\BR(\jpsi \to \mumu) = 0.0593
\pm 0.0006$ \cite{pdg} and $\BR(\Kstarz \to \Kp \pim) = 2/3$  \cite{pdg}.  The
ratio of $b$-quark hadronization factors that accounts for the different
production rate of \Bz and \Bs mesons is $f_s/f_d = 0.253 \pm
0.031$~\cite{fsfdlhcb}.  The factor 9/4 is the inverse square of the 2/3
branching fraction of $\Kstarz \to \Kp \pim$.  The number of candidate events in
the signal and control channel data samples are designated by $N_{\kstarkstar}$
and $N_{\jpsikstar}$.

The correction factor $\lambda_{f_L}$ is motivated by the fact that the overall
efficiency of the \lhcb detector is a linear function of the \Kstarz
longitudinal polarization $f_L$. Taking into account the measured value and
errors reported in section \ref{sec:Kpolarization}, Monte Carlo simulation was
used to estimate $\lambda_{f_L}=0.812\pm 0.059$.

We have considered two sources of systematic uncertainty associated to the ratio
of selection efficiencies. The first source results from discrepancies between
data and simulation in the variables related to track and vertex quality, and
the second is related to particle identification. A small difference observed in
the average impact parameter of the particles was corrected for by introducing
an additional smearing to the track parameters in the
simulation~\cite{lhcbmumu}. While the absolute efficiencies vary significantly
as a function of vertex resolution, the ratio of efficiencies remains stable. We
have assigned a $2\%$ uncertainty to the ratio, after comparison between
simulation and the \jpsikstar data. The \kaon/\pion identification efficiency
was determined using a sample of \jpsikstar events selected without making use
of the RICH detectors. As the signal channel contains one more kaon than the
control channel, a correction factor of $1.098\pm0.019$ was applied to the
branching fraction, and a 2\% error was assigned to it. The efficiency of muon
identification agrees with simulation within $1.1\%$~\cite{Jpsi_1}.  All these
factors are combined to produce an overall systematic uncertainty of 3.4\% in
the ratio of selection efficiencies. The uncertainty in the background model in
the \Bs mass fit ($\pm 2$ events) contributes an additional systematic error of
$4.7\%$.

Trigger efficiencies can be determined, for particular trigger paths in LHCb,
using the data driven algorithm described in~\cite{Jpsi_1}.  This algorithm
could be applied for the specific hadronic triggers used for \jpsikstar, but not
for the small \kstarkstar signal. The efficiency related to cuts on global event
properties, applied during the 2010 data taking, is determined from $J/\psi$
minimum bias triggers~\cite{Jpsi_1}. The result indicates a trigger efficiency
of $(26.8 \pm 3.8)\%$, smaller than the simulation result of $(31.16\pm 0.63)\%$
shown in Table~\ref{tab:BR_numbers}.  Although these are consistent within
uncertainties, we nonetheless apply a $-9$\% correction to the ratio of trigger
efficiencies between \jpsikstar and \kstarkstar channels, taking into account
correlations in the trigger probability.  A systematic error of 11\% was
assigned to uncertainty on the trigger efficiency, entirely limited by
statistics, both in the signal and control channels.  Detector occupancies,
estimated by the average number of reconstructed tracks, are larger by 10\% in
the data than in the simulation. This implies an additional correction of +4.5\%
to the ratio of efficiencies, since the control channel is observed to be more
sensitive to occupancy than the signal channel.

An $\sim$ 8\% S-wave contribution under the \Kstarz resonance in the \jpsikstar
channel  has been observed by BaBar \cite{swave-bbar}, and the data in a $\pm$70
\mevcc mass interval around the \Kstarz mass \cite{note2} yields a
($9.0\pm3.6$)\% extrapolation to the $\pm$150 \mevcc mass window.  The S-wave
background  doubles for the \Kstarz\Kstarzb final state, and it may certainly
have a  different coupling for both channels.  Our direct measurement reported
in section~\ref{sec:selyield} of (19$\pm$9)\% is still lacking precision to be
used for this purpose. When evaluating the branching fraction, we have assumed a
9\% S-wave contribution, and assigned a systematic error of 50\% to this
hypothesis. A summary of the various contributions to the systematic error can
be seen in Table \ref{tab:systematicsBR}.

\begin{table}
\centering
\caption{Estimated systematic error sources in the $\BR\left( \kstarkstar\right)$ measurement.}
\begin{tabular}{|l|c|}
\hline
Systematic effect & Error ($\%$) \\
\hline
Trigger efficiency & 11.0 \\
Global angular acceptance & 7.2 \\
S-wave fraction & 5.0 \\
Background subtraction & 4.7 \\
\begin{minipage}{1.8in} \jpsikstar and $\jpsi\to\mu\mu$ BR uncertainty \end{minipage}& 4.6 \\
Selection efficiency & 3.4 \\
\hline
Total & 15.9 \\
\hline
\end{tabular}
\label{tab:systematicsBR}
\end{table}

Our final result is
\begin{equation}
\begin{split}
\BR(\kstarkstar) = ( 2.81 & \pm 0.46\ ({\rm stat.}) \\
& \pm 0.45\ ({\rm syst.}) \\
& \pm 0.34\ (f_s/f_d) ) \times 10^{-5}. \nonumber
\end{split}
\end{equation}

As we have seen at the end of section \ref{sec:Kpolarization}, unequal
normalization factors arise upon time integration of individual polarization
amplitudes with well-defined \CP-eigenvalues.  This has the interesting
implication that the time-integrated flavor-averaged branching fraction ($B_1$)
as determined above cannot be directly compared with theoretical predictions
solely formulated in terms of the decay amplitudes
${\mathcal{A}_L}^2+{\mathcal{A}_{\parallel}}^2+{\mathcal{A}_{\perp}}^2$
($B_0$).  Meson oscillation needs to be taken into account, since two distinct
particles with different lifetimes are involved. Owing to the fact that
$\mathcal{A}_{\perp}$ is \CP-odd, the relationship between these quantities
reads as follows
\begin{equation}
B_0 = B_1  \left(1 +
\frac{\Delta\Gamma}{2\Gamma}(f_L+f_\parallel-f_{\perp}) \right).
\end{equation}
According to our measurements of $f_L+f_\parallel-f_{\perp}$, the correction is
small (3\% if current values are taken for $\Delta\Gamma$), and we do not apply
it to our measurement.

\section{Conclusion}

The $b\rightarrow s$ penguin decay $\kstarkstar$ has been observed for the first
time. Using $35$ \invpb of $pp$ collisions at 7 \tev centre-of-mass energy,
\lhcb has found $49.8 \pm 7.5 $  signal events in the mass interval $\pm 50
\mevcc$ around the \Bs mass.  Analysis of the $K^+\pi^-$ mass distributions
shows that most of the signal comes from \kstarkstar, with some S-wave
contribution.  The branching fraction has been measured, with the result
{$\BR\left( \kstarkstar\right) = (2.81 \pm 0.46 ({\rm stat.}) \pm 0.45 ({\rm
syst.}) \pm 0.34\, (f_s/f_d) )\times10^{-5}$.} The \CP-averaged longitudinal
\Kstarz polarization fraction has also been measured to be  $f_L = 0.31 \pm 0.12
({\rm stat.}) \pm 0.04 ({\rm syst.})$, as well as the \CP-odd component
$f_{\perp} = 0.38 \pm 0.11 ({\rm stat.}) \pm 0.04 ({\rm syst.})$.

When we consider our measurement in association with that of~\cite{bbar},
it is remarkable that the longitudinal polarization of the \Kstarz mesons seems
to be quite different between \kstarkstar ($f_L = 0.31 \pm 0.12 ({\rm stat.})
\pm 0.04 ({\rm syst.})$) and \Bdkstarkstar  ($f_L = {0.80}^{+0.10}_{-0.12} ({\rm
stat.}) \pm 0.06 ({\rm syst.})$), despite the fact that the two decays are
related by a U-spin rotation. However, the ratio of the branching ratios of \Bs
and \Bd decays is consistent with $1/\lambda^2$ where $\lambda$ is the
Wolfenstein parameter, as expected.



\section*{Acknowledgements}
\noindent We would like to thank J. Mat\'{\i}as for useful discussions.
We express our gratitude to our colleagues in the CERN accelerator
departments for the excellent performance of the LHC. We thank the
technical and administrative staff at CERN and at the LHCb institutes,
and acknowledge support from the National Agencies: CAPES, CNPq,
FAPERJ and FINEP (Brazil); CERN; NSFC (China); CNRS/IN2P3 (France);
BMBF, DFG, HGF and MPG (Germany); SFI (Ireland); INFN (Italy); FOM and
NWO (The Netherlands); SCSR (Poland); ANCS (Romania); MinES of Russia and
Rosatom (Russia); MICINN, XuntaGal and GENCAT (Spain); SNSF and SER
(Switzerland); NAS Ukraine (Ukraine); STFC (United Kingdom); NSF
(USA). We also acknowledge the support received from the ERC under FP7
and the Region Auvergne.




\end{document}